\documentclass[12pt,reqno]{amsart}

\usepackage[left=1in,right=1in,top=1in,bottom=1in]{geometry}
\usepackage{amsmath, amssymb, graphicx, url, mathtools}
\usepackage[round]{natbib}
\usepackage{xcolor}
\usepackage{wrapfig}
\usepackage{tikz}
\usepackage{mathptmx}

\usepackage{tikz}
\usepackage{multirow}
\usepackage{mathtools} 
\usepackage{wrapfig}
\usepackage{bm}
\usepackage{paralist}
\usepackage{accents}
\usepackage{enumerate}
\usepackage{caption}

\usepackage{mathptmx}

\usepackage{amsmath,amssymb,amsthm}
\usepackage{url,mathtools}
\usepackage{colortbl,framed}
\usepackage{enumitem}

\usepackage{float}
\usepackage[round]{natbib}
\usepackage{placeins}

\theoremstyle{definition}

\numberwithin{remark}{section}

\numberwithin{equation}{section}
\numberwithin{theorem}{section}

\usepackage{url}
\newcommand{\citen}{\citet}

\usepackage{pagecolor,lipsum}

\usepackage{wallpaper}
\usepackage{graphicx}

\usepackage{xr}
\makeatletter

\newcommand*{\addFileDependency}[1]{
\typeout{(#1)}
\@addtofilelist{#1}
\IfFileExists{#1}{}{\typeout{No file #1.}}
}\makeatother

\newcommand*{\myexternaldocument}[1]{%
\externaldocument{#1}%
\addFileDependency{#1.tex}%
\addFileDependency{#1.aux}%
}
\myexternaldocument{Online_Appendix}

\begin{document}

\begin{center}
\bigskip

\end{center}

\title[Generic ML for Features of Heterogenous Treatment Effects]{Reply to: comments on ``Fisher-Schultz Lecture: Generic Machine Learning  Inference on Heterogenous Treatment Effects in Randomized Experiments, with an Application to 
Immunization in India''}
\author{Victor  Chernozhukov \and  Mert Demirer \and Esther Duflo  \and Iv\'an Fern\'andez-Val}\thanks{The authors respectively from MIT, MIT, MIT and BU}
\date{\today}
\maketitle

We warmly thank Kosuke Imai, Michael Lingzhi Li, and Stefan Wager for their gracious and insightful comments. We are particularly encouraged that both pieces recognize the importance of the research agenda the lecture laid out, which we see as critical for applied researchers. It is also great to see that both underscore the potential of the basic approach we propose—targeting summary features of the CATE after proxy estimation with sample splitting.

We are also happy that both papers push us (and the reader) to continue thinking about the inference problem associated with sample splitting. We recognize that our current paper is only scratching the surface of this interesting agenda. Our proposal is certainly not the only option, and it is exciting that both papers provide and assess alternatives. Hopefully, this will generate even more work in this area.

\section{Response to \citen{wager2024sequentialvalidationtreatmentheterogeneity} }

One potential concern with our approach is that it is demanding in terms of data, since it relies on repeated splitting of data into two parts: one used for CATE signal extraction and another used for post-processing. To examine potential improvements, Wager's discussion focuses on the special problem of testing the null effect—that is, whether the CATE function is zero. It  is a specific setting, as typical machine learning algorithms are in fact able to learn the zero function consistently even in high-dimensional settings.\footnote{This happens because the typical algorithms penalize deviations away from the zero function.} Nonetheless, the problem of testing the null of a zero-CATE remains very important.

Fixing a \textit{single} split of data into $K$ folds, \citen{wager2024sequentialvalidationtreatmentheterogeneity} investigates relative gains in power generated by the sequential inference approach of \citet{luedtke2016statistical}. This approach uses progressively more data to estimate the ``signal" and then generates a sequence of statistics to test if the ``signal" is zero. The statistics can be aggregated to form a ``single-split" p-value using the martingale properties of the construction. Wager shows in Monte Carlo experiments (reproduced below) that this improves power over a method of taking the median p-value over $K$ equal-sized folds (which is not the method we propose, but is a sensible  benchmark). It also outperforms the ``naïve" approach that relies on cross-fitting à-la debiased machine learning (DML) approach, which is asymptotically valid in this special setting but suffers from size distortions.\footnote{Under the null CATE hypothesis, DML provides an asymptotically valid testing approach for any generic ML method that can consistently estimate the zero function in a given high-dimensional setting.}

We believe that Wager's proposal is potentially a fruitful complement to what we propose, since we can use the sequential estimation within our ``multiple-split" approach. We now show that this combination generates further size and power improvements.

In what follows, we report results from a numerical simulation using the same experiment and implementation details as in \citen{wager2024sequentialvalidationtreatmentheterogeneity}.\footnote{We are very grateful to Stefan Wager for sharing the code of \citen{wager2024sequentialvalidationtreatmentheterogeneity} with us.} As in Wager, we consider the following ``single-split" approaches: (a) naïve or DML approach; (b) 2-fold approach with 2/3 of the sample allocated to training and 1/3 to testing data; and (c) the sequential approach (with 3 equal folds). We compare these approaches to "multiple-split" versions of (a), (b), and (c).

The results in Table \ref{table:t2}, based on 10,000 simulation replications, show that, in line with Wager, the sequential approach increases power relative to the simple approach of using two folds of unequal size. Interestingly, the use of "multiple" splits makes the sequential approach even better: the frequency of false rejection is decreased dramatically while the power of rejecting the false null increases.

Finally, recall that we are testing here the null CATE. ``Multiple" splitting fixes the size distortions of the ``naïve" DML method, and it emerges as a very strong winner among all—it has the highest power and keeps the size well below the nominal level. Of course, we don't expect this superior performance of the naïve method to hold in more general settings of inference on CATE features, whenever the CATE function is not "special" enough to be learned quickly by ML (zero function, flat function, approximately sparse, etc).

\begin{table}[ht]
\centering
\caption{Tests of Zero CATE with 5\% Significance Level}\label{table:t1}
\begin{tabular}{lrrrrrr}
\hline\hline
& \multicolumn{3}{c}{Single Split: Wager} & \multicolumn{3}{c}{Multi (100) Splits: CDDF} \\
& ``Naïve" & 2-Folds & Sequential & ``Naïve" & 2-Folds & Sequential \\ 
\hline
Size (False Rejection) & 8.40\% & 5.11\% & 4.65\% & 0.81\% & 0.00\% & 0.03\% \\ 
  Power (Correct Rejection) & 81.01\% & 44.73\% & 62.71\% & 84.25\% & 41.11\% & 64.68\% \\ 
\hline\hline
\end{tabular}
\end{table}

{\footnotesize \textit{Notes: 10,000 simulation replications, sample size is 1,000. Size obtained for CATE function $\tau(z) = 0$ and power for $\tau(z) = (z_1)_+$. }}

In summary, we believe that the idea of \citen{wager2024sequentialvalidationtreatmentheterogeneity} of using martingale aggregation warrants further investigation, and we very much welcome any further research in this area.\footnote{The theory in \citet{luedtke2016statistical} seems to rely on consistent learning, but this can probably be extended to pseudo-consistent learning. This extension can potentially cover some ground in ML applications in high-dimensional settings.}

\section{Response to \citen{imai23}}

\citen{imai23} propose an alternative inference approach to account for sample-splitting uncertainty, based on Neyman's randomization paradigm. Relative to our approach, their method is analytical and relies on a single split of the data. This provides a clear computational advantage, as performing multiple splits requires additional computation time.\footnote{In its current form, their method is tailored to GATES, although \citen{imai23} suggest it could be modified for other purposes.} We find the approach very interesting, although we have two comments. 

First, as we highlighted theoretically in our paper, our multiple-split approach outperforms the single-split approach in terms of estimation risk. Specifically, we formally established that our method has a lower mean absolute deviation (MAD). We provide empirical evidence for this claim in the computational experiment below.

Second, as emphasized in our paper, a key motivation for our approach is its natural protection against ``data mining'' (whether intentional or not). For example, a researcher might try a few ($F$) different Monte Carlo seeds and ---for replicability purposes--- retain the seed that produces the most favorable results. This ``mining'' behavior in single-split approaches significantly increases estimation risk. In contrast, our procedure is expected to remain highly stable and exhibits minimal to no distortion. We provide practical evidence for this point in the computational experiment below.

In Table \ref{table:t2}, we reuse the Monte Carlo design from the previous section, where the CATE is zero, $\tau(z) = 0$. The table reports the bias, standard deviation, and MAD of estimators for the difference in GATES and rejection frequencies for this parameter being equal to zero, based on single and multiple sample splits. Specifically, we compare the method of \citen{imai23}, which uses three folds ($L=3$) with cross-fitting and a single split (IMLI), to our method (CDDF), computed as the median of 100 splits, with $2/3$ of the sample in the auxiliary set and $1/3$ in the validation set.\footnote{The method of \citen{imai23} is implemented using the R package \texttt{evalITR} \citep{evalITR}.} The columns labeled ``Mining ($F=5$)'' illustrate the risks of data mining when using estimators reliant on a single split of the data. These columns report results for the maximum of IMLI and CDDF over $F = 5$ different random seeds, emulating the behavior of a ``mining'' researcher searching (intentionally or not) for positive effects.

From Table \ref{table:t2}, we draw the following conclusions:

\begin{enumerate}
    \item \textbf{Estimation Risk:} CDDF exhibits a much smaller root-mean-square error and MAD than IMLI, consistent with the theoretical advantage of CDDF over single-split procedures.
    \item \textbf{Stability Against Mining:} Mining the IMLI procedure significantly increases the risk of estimation errors, whereas CDDF remains relatively stable and insensitive to mining.
    \item \textbf{Conservativeness:} Both IMLI and CDDF are conservative inferential procedures, with false rejection rates (size) substantially below the nominal $5\%$ level. They remain conservative even under a modest degree of mining. However, CDDF demonstrates a lower false rejection rate while maintaining smaller estimation risk, indicating more statistically desirable properties overall.
\end{enumerate}

\begin{table}[ht]
\centering\caption{Finite-Sample Properties}\label{table:t2}
\begin{tabular}{lrrrr}
  \hline\hline
 & \multicolumn{2}{c}{ } & \multicolumn{2}{c}{Mining ($F=5$)} \\
 & IMLI & CDDF & IMLI & CDDF \\ 
  \hline
Bias & 0.00 & 0.00 & 0.14 & 0.03 \\ 
  SD & 0.15 & 0.09 & 0.12 & 0.09 \\ 
  MAD & 0.10 & 0.06 & 0.13 & 0.06 \\ 
  Size  (False Rejection) & 0.40\% & 0.00\% & 1.66\% & 0.01\% \\ 
  \hline\hline
\end{tabular}
\end{table}

{\footnotesize \textit{Notes: $10,000$ simulation replications, sample size is $1,000$, and $\tau(z) = 0$. The parameter is the difference in GATES with two groups. IMLI is the estimator of \citen{imai23} with 3 folds. CDDF is our estimator, computed as the median of $100$ splits, with $2/3$ of the sample in the auxiliary set and $1/3$ in the validation set.  Mining ($F=5$) computes the maximum estimator over $5$ different random seeds.}}

In summary, we conclude that performing multiple splits provides clear statistical advantages over single-split methods, provided the computational cost is not a significant concern. CDDF offers lower estimation risk, greater robustness to mining, and attractive inferential properties.

\section{Concluding Remarks}
Developing reliable  methods to uncover the presence and magnitude of heterogeneous treatment effects is an important task in modern econometrics and statistics. Our paper made a specific suggestion, and  \citen{wager2024sequentialvalidationtreatmentheterogeneity} and  \citen{imai23} propose clever alternatives, which are computationally appealing because they do not require multiple splits. Unsurprisingly, these gains come with some costs both in terms of theoretical requirements, and potential robustness to data mining. We see these approaches are very useful complements to the idea of multiple splits. Additional research on how to balance these trade-off would be highly valuable.

\bibliographystyle{plainnat}
\bibliography{mybibVOLUME.bib}

\end{document}